\documentclass[prd,twocolumn,showpacs,preprintnumbers]{revtex4}
\pdfoutput=1
\usepackage[plainpages=false, colorlinks=true, anchorcolor=blue, linkcolor=blue, citecolor=blue, bookmarks=false]{hyperref}
\usepackage{amsfonts,amsmath,amssymb}
\usepackage{graphicx}
\usepackage{color}
\usepackage{natbib}
\graphicspath{c:/Users/landau/Downloads}
\pdfoutput=1
\newcommand{\rthis}[1]{\textcolor{black}{#1}}

\begin{document}

\newcommand{\apjl}{Astrophys. J. Lett.}
\newcommand{\apjs}{Astrophys. J. Suppl. Ser.}
\newcommand{\aap}{Astron. \& Astrophys.}
\newcommand{\aj}{Astron. J.}
\newcommand{\araa}{Ann. Rev. Astron. Astrophys. } 
\newcommand{\mnras}{Mon. Not. R. Astron. Soc.}
\newcommand{\jcap}{JCAP}
\newcommand{\pasj}{PASJ}
\newcommand{\pasa}{Pub. Astro. Soc. Aust.}
\newcommand{\pasp}{Pub. Astro. Soc. Pacific}
\newcommand{\physrep}{Physics Reports.}

\title{Constraints on differential Shapiro delay  between  neutrinos and photons from IceCube-170922A}
\author{Sibel \surname{Boran}$^1$}\altaffiliation{E-mail: borans@itu.edu.tr}
\author{Shantanu  \surname{Desai}$^2$} \altaffiliation{E-mail: shantanud@iith.ac.in}
\author{Emre  O. \surname{Kahya}$^3$} \altaffiliation{E-mail: eokahya@itu.edu.tr}

\affiliation{$^{1,3}$Department of Physics, Istanbul Technical University, Maslak 34469 Istanbul, Turkey}
\affiliation{$^{2}$Department of Physics, Indian Institute of Technology, Hyderabad, Telangana-502285, India}

\begin{abstract}

 On 22nd September 2017, the IceCube Collaboration detected a neutrino with energy of about 290 TeV from the direction of  the gamma-ray blazar TXS 0506+056, located at a distance of about 1.75 Gpc. During the same time, enhanced gamma-ray flaring was also simultaneously observed from multiple telescopes,  giving rise to only the second coincident astrophysical neutrino/photon observation after SN 1987A. We point out that for this event,  both neutrinos and photons encountered a  Shapiro delay of about {\color{black}6300} days along the way from the source.  From this delay and the relative time difference   between the neutrino and photon arrival times, one can  constrain violations of  Einstein's Weak Equivalence Principle (WEP) for TeV neutrinos. We constrain   such violations of WEP  using the Parameterized Post-Newtonian (PPN) parameter $\gamma$, which is given by  $|\gamma_{\rm {\nu}}-\gamma_{\rm{EM}}|<5.5 \times 10^{-2}$, after assuming  time difference of \rthis{175} days between neutrino and photon arrival times. 
\pacs{97.60.Jd, 04.80.Cc, 95.30.Sf}
\end{abstract}
\maketitle

\section{Introduction}

On 22nd September 2017,  the IceCube experiment detected a \textcolor{black}{neutrino-induced upward-going} muon (called IceCube-170922A) having an energy of about 24 TeV, \textcolor{black}{ with a 56\% chance of been produced by an astrophysical neutrino. Its parent neutrino energy has been estimated to be  about  290 TeV~\citep{IceCube}.} The location of the neutrino direction was found to be in positional coincidence with a TeV gamma-ray blazar. Following this detection, the  Fermi-LAT and AGILE gamma-ray satellites reported enhanced flaring from this blazar during this time period, in the energy range between   0.1 to  10 GeV. About a week after the IceCube neutrino detection, the MAGIC atmospheric Cherenkov telescope  also reported a 6$\sigma$ excess of 100 GeV gamma-rays compared to the expected background. The probability that the association of IceCube-170922A with the gamma-ray flare is a chance coincidence is disfavored at 3$\sigma$ \textcolor{black}{after considering three models of correlated neutrino/gamma-ray emission}. More details about these observations can be found in Ref.~\citep{IceCube}.
This is the first direct evidence that blazars  produce high-energy neutrinos and probably ultra-high-energy cosmic-rays.

These observations also enable a novel probe of General Relativity (GR) and the equivalence principle for TeV neutrinos. The total time it takes for neutrinos  to reach the Earth from this blazar is equal to the sum of the distance divided by their (vacuum) speed  and an additional delay due to the non-zero gravitational potential of the cumulative mass distribution along the line-of-sight.

The latter delay is known as Shapiro delay~\cite{Shapiro} and has been extensively used to test GR as well for astrophysical measurements for the past five decades~\cite{Shapiro66,Cassini,Kopeikin,Will,Taylor94,Demorest}.
The first ever calculation of this  line-of-sight cumulative Shapiro delay was done in 1988, following the detection of neutrinos from SN~1987A at a distance of 50 kpc~\cite{IMB,Kamioka} and optical light soon after the neutrino event. After this detection, two groups~\cite{Longo,Krauss} (see also Ref.~\citep{Franson}) pointed out that the neutrinos underwent  a Shapiro delay of about 1-6 months due to the gravitational potential of the Milky Way (MW) galaxy. We note that until the  current IceCube detection, this was the only direct  and ``smoking gun" evidence  that neutrinos are affected by GR and obey WEP to a precision of 0.2-0.5\%. Since IMB and Kamiokande detected both neutrinos and anti-neutrinos, these observations also  showed that Shapiro delay for neutrinos is CP invariant~\citep{LoSecco}. Subsequently,  two years ago, based on a 2$\sigma$ putative association between a giant flare from the blazar PKS B1424-418 located at $z=1.522$, and a 2 PeV IceCube neutrino, WEP was constrained to an accuracy of $10^{-5}$~\citep{Wang16}. We also note that  in Ref.~\cite{weijcap}, based on associations between five  cascade neutrino events detected by IceCube and GRBs located at redshifts  greater than 0.1, WEP was constrained to a much higher value $\mathcal{O} (10^{-11} -10^{-13})$. However,
the $p$-value for this coincidence is about 0.32 and these neutrino events are consistent with background~\cite{weijcap}.

Since 2016, there has been a proliferation of papers carrying out  similar calculations of line-of-sight Shapiro delay for a large class of astrophysical objects~\cite{grb,Wei,Kaplan,Wu17,blazar,Nusser,Kahya16,Desai17,Boran18,Murase,LVCFermi,Meszaros17,Fan,Bertolami} (and references therein).  The main objective of these works was to constrain WEP using  coincident electromagnetic (EM) wave or gravitational wave observations. The violation of WEP in these works has been parameterized in terms of the difference in  PPN parameter  $\Delta \gamma$~\cite{Will} between the astrophysical messengers.  \textcolor{black}{A summary of previous results using EM observations can be found in Table I of Ref.~\cite{Wu17}. The corresponding bounds on PPN $\gamma$ using photons range from $\mathcal{O}(10^{-3})$ to $\mathcal{O}(10^{-16})$~\cite{Wu17}. From gravitational wave observations of the first binary neutron star merger, $\Delta \gamma$ between gravitational waves and photons was constrained to be less than $\mathcal{O}(10^{-8})$~\citep{Desai17}. The limits on difference in PPN $\gamma$ between neutrinos and photons from other works (along with our result) are summarized in Table~\ref{table}.}

Now that the first simultaneous detection of TeV neutrinos and high energy photons has happened, we carry out a similar test of WEP  for neutrinos, by first calculating the line-of-sight Shapiro delay and then using the observed time difference between the neutrinos and gamma-rays to constrain the violation of WEP in terms of $\Delta \gamma$. This bound will be the first ever limit for TeV neutrinos, since the bounds obtained in Refs.~\cite{Longo,Krauss} are in MeV energy range \textcolor{black}{and the previous results for TeV neutrinos and above~\citep{Wang16,weijcap} were based upon marginal detections.} 


\section{Shapiro Delay calculation}
\label{sec:dmemulators}

The location of the neutrino direction was traced to RA$=77.43^{\circ}$, and $\delta=+5.69^{\circ}$~\citep{IceCube}. This was found to be in positional coincidence with a TeV gamma-ray blazar (TXS 0506+056). Its redshift was estimated to be $0.3365 \pm 0.001$, from high resolution optical spectroscopy in the wavelength range from 4100-9000 \AA, with the 10.4 m telescope in Canary Islands~\cite{redshift}. This redshift corresponds to a luminosity distance of approximately 1.75~Gpc, obtained using the online Cosmology  calculator~\citep{Wright} (assuming the cosmological parameters as $H_{0}=69.6$ km s$^{-1}$ Mpc$^{-1}$, $\Omega_{m}=0.286$ and $\Omega_{\Lambda}=0.714$). 

To  perform a ball-park estimate of  the Shapiro delay calculation, we follow the same procedure as in Ref.~\citep{Kahya16}, which was used to obtain the delay for GW150914. We also note that the non-zero mass of the neutrino does not change the Shapiro delay as compared to a massless carrier~\cite{Bose}, and we can calculate this delay using GR.
We first use  the total estimated Shapiro delay for our Milky-Way (MW) galaxy by using the Navarro-Frenk-White (NFW) dark matter profile~\citep{NFW}  from our previous works; this MW-induced delay is about 300 days at a distance of 400 kpc~\cite{Kahya10,Desai}.
Once we go beyond the galaxy virial radius, the delay follows a logarithmic  trend as a function of distance for a point-like source and Schwarzschild metric:
\begin{equation}
\Delta t_{\rm{shapiro}}=(1+\gamma) \frac{GM}{c^3} \ln\left(\frac{d}{b}\right)\; ,
\end{equation}
\noindent where $\gamma$ is the PPN parameter with value equal to unity in GR, $b$ is the impact parameter, and $d$ is the distance to the source. To calculate the total Shapiro delay due to our galaxy  at a distance of {\color{black}1.75}~Gpc, we assume $b$=8~kpc and use the logarithmic enhancement factor from the above equation, which is equal to about a factor of three. Therefore,
the Shapiro delay due to only our MW galaxy at a distance of  {\color{black}1.75}~Gpc is equal to 900 days.  

Next, we estimate the total number of MW equivalent galaxies by considering a cylindrical line-of-sight, whose surface area  is determined by the galaxy virial radius and height by the distance to the source. The same procedure was used to estimate the extra-galactic Shapiro delay  contribution to GW150914~\citep{Kahya16}. This extra factor is given by $(r_{vir}/d)^2 \times N_{tot}$; where $r_{vir}$ is the virial radius, $d$ is the distance to the source and $N_{tot}$ is the total number of MW equivalent galaxies, 
{\color{black} $N_{tot}$ equal to} ${\color{black}\sim 3.49} \times 10^{8}$ at a distance of {\color{black}1.75}~Gpc~\citep{Simhil}. 
This estimate is  comparable with that in Ref.~\cite{Conselice}, given by  $N_{tot}= 3.1\times10^{8}$ at this redshift.
Using the  estimate of $N_{tot}$  from Ref.~\citep{Simhil}  and  $r_{vir}=250$ kpc, we get a total of {\color{black} 7} MW like galaxies in our cylindrical tube. Therefore, the total estimated Shapiro delay is equal to about {\color{black}6300} days. We note that in order to detect these seven galaxies, one would need high quality deep multi-band imaging  data around this blazar in order to infer the photometric redshifts and masses of these galaxies using Spectral Energy Distribution (SED) fitting. This is not possible using the available spectroscopic data for this blazar~\cite{redshift}. To the best of our knowledge only deep $V$-band data is available for this blazar, from which it is not possible to detect these seven MW-like galaxies.

\rthis{We note that the above delay of 6300 days  is a conservative estimate, and does not include the contribution of lower mass galaxies. To get a ballmark estimate of this contribution, we can use the compilation of galaxies from the Gravitatonal Wave Galaxy Catalog (GWGC)~\cite{daw}, which is valid upto 100 Mpc to get an estimate of the fraction  of low mass 
galaxies, and assume that this  fraction is approximately the  same at a distance of 1.75 Gpc at the location of TXS 0506+056. The GWGC provides $B$-band magnitude for all the galaxies, which can serve as a proxy for the total mass~\cite{peebles}. The ratio of the total number of galaxies fainter than the MW B-band magnitude of -20.6, to those brighter than this magnitude is about 11. Therefore we expect about 77 galaxies with masses lower than MW along the line of  sight to the blazar, given our estimate of 7 MW-equivalent galaxies. However, the total mass contribution from these low mass galaxies is about 10\% of the total contribution, compared to the MW-equivalent galaxies in this catalog. Assuming we can extrapolate this ratio from 100 Mpc to 1.75 Gpc, the Shapiro delay would be underestimated by about 10\%, since it scales directly with the mass.} Furthermore, we have also neglected
the effects due to Hubble expansion, which would enhance the Shapiro delay estimates computed here~\cite{Nusser}.



\section{Constraints on WEP}
\label{sec:WEP}
 Once the Shapiro delay for a given mass distribution is known,  if the neutrinos and gamma-rays arrive from the same source  within a time interval equal to $\Delta t$, one can constrain the violations of WEP in terms of the PPN parameter $\Delta \gamma \;=\; | \gamma_{\rm{\nu}} -\gamma_{\rm{EM}} |$ and the calculated Shapiro delay $\Delta t_{\rm{shapiro}}$ using~\cite{Wei}:
\begin{equation}
\Delta\gamma \leq 2 \; \frac{\Delta t}{\Delta t_{\rm {shapiro}}}\;.
\label{eq:1}
\end{equation}
 In Ref.~\cite{IceCube}, data for enhanced gamma-ray emission from the Fermi-LAT  around the time of the IceCube neutrino event has been presented, after binning the data into seven-day intervals. The peak of the high energy gamma-ray emission happened  15 days prior to the IceCube signal~\cite{IceCube}. \textcolor{black}{To evaluate the chance probabilities of this association, three different models for association between the neutrino and gamma-ray flux were considered. In the first model, the neutrino flux is linearly correlated  with the gamma-ray flux. In the second model, the neutrino flux is strongly correlated with variations in gamma-ray flux. In the third model, a correlation is assumed between the neutrino and TeV gamma-ray flux}. In all these models,  the  probability that this  temporal association is a chance coincidence can been rejected at 3$\sigma$, after accounting for the ``look elsewhere'' effect~\citep{IceCube}.  \rthis{Hence, we cannot definitely say that the IceCube-170922A neutrino is associated with photons observed during the peak of the gamma ray emission. The maximum possible time difference between the start of the flare and the neutrino is about 175 days. Therefore, for a conservative estimate we use $\Delta t =175$ days, to obtain a bound on WEP.}

Therefore, assuming \rthis{$\Delta t=175$ days, we get $|\Delta \gamma| < 5.5 \times 10^{-2}$} from Eq~\ref{eq:1}.  This is the first test  of WEP for TeV neutrinos over a distance of {\color{black}1.75}~Gpc, as the previous multi-messenger astrophysical neutrino detection occurred in the MeV energy range and at  a distance of 50 kpc. We  note that this limit is conservative. For a more robust estimate, the difference in time between neutrino and photon production at the source must be subtracted from the observed arrival time differences. \rthis{If we assume that the arrival time difference corresponds to  the difference between the peak of  gamma-ray emission and IceCube neutrino signal ($\Delta t$ = 15 days), then  $|\Delta \gamma| < 4.7 \times 10^{-3}$}.
However, such a  limit would only be  valid  within the context of models,  which predict  neutrino emission in a narrowly confined time window around the very peak of the gamma-ray emission.

 We note that the IceCube Collaboration has also searched 9.5 years of data starting from 2008 to look for excess emission of high energy neutrinos and found a 3.5$\sigma$ excess between Sept. 2014 and Mar. 2015~\cite{IceCubepre}. If the detection of IceCube-170922A is due to  a steady-state  emission mechanism rather than  a flaring emission, the time window can be as large as 9 years, corresponding to $\Delta \gamma < 1.03$. For any other time window, the limit on  $\Delta \gamma$ can be scaled according to the time window used.

\section{Conclusions}
\label{sec:concl}

After the announcement of the first concurrent detection of TeV neutrinos and gamma-ray photons by the IceCube and multiple gamma-ray collaborations~\cite{IceCube}, we calculated the line-of-sight Shapiro delay to this source (IceCube-170922A) following  our previous works~\cite{Desai,Kahya08,Kahya10,Desai15,Kahya16,Desai17,Boran18}. The total estimated Shapiro delay is about \textcolor{black}{6300} days. The coincident detection of  gamma-rays, allows us to test WEP for neutrinos. Assuming the neutrinos and photons arrived within an  interval of \rthis{175} days , one can constrain WEP from  the difference in PPN $\gamma$ parameters between the neutrinos and photons, and is given by   $|\gamma_{\rm{\nu}} - \gamma_{\rm{EM}}|< 5.5\times 10^{-2}$. \rthis{For any other time window between neutrino and photons, the limit on PPN $\gamma$ can be scaled accordingly.} This is the first such test for TeV neutrinos and is complementary to the limit obtained for MeV neutrinos from SN 1987A~\citep{Longo,Krauss}. 

{\bf Note Added:} After this work appeared on arXiv, we found two other works~\citep{Laha,Wei18}, which have also  independently set a limit on WEP for this event. Both of these works \textcolor{black}{modeled the gravitational potential of the Laniakea supercluster as a point source to estimate  the Shapiro delay. Therefore, this calculation is complementary to ours. We note that there is still considerable uncertainty regarding the mass and gravitational potential of the Laniakea  supercluster~\cite{weijcap}.} Using $\Delta t=7$ days and  $\Delta t=15 (175)$  days, the inferred $\Delta \gamma$ values in these works are given by $|\Delta \gamma| < 3.5 \times 10^{-7}$~\cite{Laha} and $|\Delta \gamma| < 7.3 \times 10^{-7}$ ($8.5\times 10^{-6}$)~\citep{Wei18}, respectively. 

{ \color{black} In order to compare our result to these studies as well as with previous results from SN1987A, we have tabulated these results in Table~\ref{table}.

\newpage

\begin{widetext}

\begin{table}[htbp]
\begin{center}
\caption{The limits on violation of WEP between neutrinos and photons from previous literature,  along with the result from this work.}
\begin{ruledtabular}
\begin{tabular*}{1.5\textwidth}{lllll} 
$\rm Source$&$\rm Messengers$&$\rm Gravitational\;Field$&$\Delta t = |t_{\nu} - t_{\rm EM} | $&$ \Delta \gamma = |\gamma_{\nu} - \gamma_{\rm EM }|$ \\[1ex]
\hline\\
$\rm SN \;1987A$&$ \rm \nu(MeV) - EM (eV)$ & $\rm Milky\;Way$& $ 6\; {\rm hrs}$ \cite{Longo} &$ 3.4 \times 10^{-3}$ \cite{Longo}\\[2ex]

$\rm SN \;1987A$&$ \rm \nu(MeV) - EM (eV)$ & $\rm Milky\;Way$& $ 10^4 \; {\rm s}$ \cite{Krauss} &$4\times 10^{-3}\;{\rm and}\; 7\times 10^{-4}$ \cite{Krauss}\\[2ex]

$\rm Blazar\;TXS\;0506+056$&$ \rm \nu(PeV) - EM (eV)$ & $\rm Laniakea\;supercluster$ & $ 7 \; {\rm days}$ \cite{Laha} &$ 3.5 \times 10^{-7}$ \cite{Laha}\\[2ex]
& & $\rm of\;galaxies$ &  &\\[2ex]

$\rm Blazar\;TXS\;0506+056$&$ \rm \nu(TeV) - EM (eV)$ & $\rm Laniakea\;supercluster$ & $15 \; {\rm days}$ \cite{Wei18}&$ 7.3 \times 10^{-7}$ \cite{Wei18}\\[2ex]
& & $\rm of\;galaxies$ &  &\\[2ex]

$\rm Blazar\;TXS\;0506+056$&$ \rm \nu(TeV) - EM (eV)$ & $\rm Laniakea\;supercluster$ & $ 175 \; {\rm days}$ \cite{Wei18}&$ 8.5 \times 10^{-6}$ \cite{Wei18}\\[2ex]
& & $\rm of\;galaxies$ &  &\\[2ex]

$\rm Blazar\;TXS\;0506+056$&$ \rm \nu(TeV) - EM (eV)$ &  $\rm All\;Milky\;Way-like$ & $  175 \; {\rm days \;}$ &$5.5 \times 10^{-2}\;{\rm (This \;work)}$\\[2ex]
& &  $\rm galaxies\;along\;line-of-sight$  &  &\\[2ex]

\end{tabular*}\label{table}
\end{ruledtabular}
\end{center}
\end{table}

\end{widetext}
}


\begin{acknowledgements}
E.O.K. acknowledges support from TUBA-GEBIP 2016, the Young Scientists Award Program. We are grateful to Richard Woodard for prior collaboration on the Shapiro delay  calculations.
\end{acknowledgements}

\bibliography{dmemulator}
\end{document}